\documentclass[aps,pra,notitlepage,twocolumn]{revtex4-1}
\usepackage{amsmath,amssymb}
\usepackage{dsfont}
\usepackage[utf8]{inputenc}
\usepackage[T1]{fontenc}
\usepackage{graphicx}
\usepackage[usenames, dvipsnames]{color}
\usepackage{hyperref}

\begin{document}

\newcommand{\bra}[1]  {\langle #1|}
\newcommand{\ket}[1]  {| #1\rangle}
\newcommand{\braket}[2]  {\langle #1|#2\rangle}
\newcommand{\av}[1]  {\langle #1 \rangle}

\title{Quantum-enhanced interferometry with cavity-QED-generated non-classical light}

\author{Karol Gietka, Tomek Wasak, Jan Chwede\'nczuk}
\affiliation{Faculty of Physics, University of Warsaw, ul. Pasteura 5, 02--093 Warsaw, Poland}
\author{Francesco Piazza, Helmut Ritsch}
\affiliation{Institut f{\"u}r Theoretische Physik, Universit{\"a}t Innsbruck, 
Technikerstra\ss{}e 21, A-6020 Innsbruck, Austria}

\begin{abstract}
We propose an enhanced optical interferometer based on tailored non-classical light generated 
  by nonlinear dynamics and projective measurements in a three-level atom cavity QED system. A coherent state
  in the cavity becomes dynamically entangled with two ground
  states of the atom and is transformed to a macroscopic superposition state via a projective measurement on the atom.
  We show that the resulting highly non-classical state can improve interferometric precision measurements 
  well beyond the shot-noise limit once combined with a classical laser pulse at the input of a
  Mach-Zehnder interferometer. For a practical implementation, we identify an efficient phase shift estimation 
  scheme based on the counting of photons at the interferometer output.
  Photon losses and photon-counting errors deteriorate the interferometer sensitivity, 
  but we demonstrate that it still can be significantly better than the shot-noise limit under realistic conditions.
  \end{abstract}

\maketitle

\section{Introduction}
Measurement is a process of transferring information from an object under investigation to the detector, 
be it a ruler, a radio telescope, or a human eye.
For the measurement to be precise, the detector must be susceptible to small perturbations induced by 
the object. This susceptibility can be quantified by $\Delta\theta$---the best
resolution attainable during the measurement of a physical quantity
$\theta$.
With $N$ being the number of intereferometric resources i.e. the
number of entities in which information about $\theta$ is encoded
during the measurement, and in absence of quantum correlations between
those entities, there is a lower bound to the value of $\Delta\theta$, namely the shot-noise limit (SNL). It states that $\Delta\theta\geqslant\frac1{\sqrt N}$ \cite{carmichael1987spectrum}.
Overcoming the SNL can become crucial for the detection of small
perturbations such as minuscule variations in the gravitational
acceleration experienced by an ultra-cold coherent Bose gas
\cite{kasevich,robins} or the detection of the gravitational waves \cite{mckenzie2002experimental,waves1,waves2}. 

To improve beyond the SNL, one must introduce tailored quantum correlations to the detector system \cite{pezze2009entanglement,giovannetti2004quantum}.
Quantum states can be more susceptible to perturbations than the classical ones and, in principle, the resolution $\Delta\theta=\frac1{N}$ can be achieved.
While this Heisenberg-limited sensitivity was considered a largely academic
concept, major progress has recently been made in the preparation of many-body entangled states potentially useful for 
ultra-precise quantum metrology. This goal was achieved by manipulating a highly-controllable
and coherent Bose gas to create spin-squeezed states \cite{esteve2008squeezing,appel2009mesoscopic,gross2010nonlinear,riedel2010atom,leroux2010orientation,chen2011conditional,berrada2013integrated,smerzi_ob} or systems with correlated atomic pairs \cite{cauchy_paris,collision_paris,twin_paris,twin_beam}.

Light interferometers also improve their performance when operating on non-classical electromagnetic fields \cite{glaub,sud}, for instance, formed by 
pairs of photons obtained in the parametric down conversion \cite{pdc1,pdc2,jachura2016mode}.
The Mach-Zehnder interferometer (MZI) fed with a coherent beam through one port and a squeezed vacuum through the other one 
yields the sub-shot-noise (SSN) scaling of $\Delta\theta$
with the intensity of light \cite{pezze_mzi,Pezze2009PRLMZI}.

Here we discuss a quantum non-demolition (QND) protocol \cite{brune1992manipulation,ritsch_1997,ritsch_2004,duan_2005} involving an optical cavity
mode interacting with a single atom. We demonstrate that this protocol yields a non-classical state of light \cite{asboth2005computable} that
combined with a coherent pulse at the input of an optical interferometer
can provide SSN sensitivity of the phase estimation. This result
persists in the presence of realistic cavity losses and moderate
imperfections of the photon counting at the output of the
interferometer. Our protocol provides an alternative to parametric
down conversion, allowing for larger entanglement for applications
where low intensities are desirable.

A successful QND protocol generating non-classical light has been introduced at ENS Paris
\cite{brune1992manipulation} using the superconducting microwave cavities. Similar proposals followed in the optical regime \cite{ritsch_1997,PhysRevA.58.3472,PhysRevA.59.4095,ritsch_2004,duan_2005}
where the generated field emitted from the resonator can be directly analyzed or used for intereferometric
protocols, as studied in the following. Recent technical progress in microwave amplification and detection also allows for direct microwave photon detection in the co-planar waveguide cavities
\cite{peropadre2011approaching}.

The manuscript is organized as follows. First, we describe the method of generating non-classical state of light via the QND protocol. Next, we introduce the intereferometric scheme and
characterize its sensitivity. Subsequently, we introduce two distinct sources of noise to the interferometric scheme. Finally, we provide the concluding remarks.

\section{Generation of a non-classical cavity field via atom-based QND measurements}

\begin{figure}[t!]
  \centering
  \resizebox{0.3\textwidth}{!}{%
  \includegraphics[]{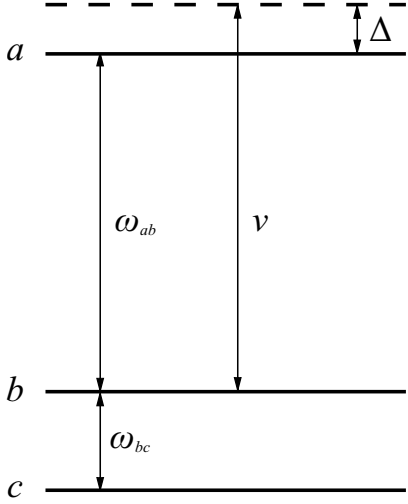}

}
  \caption{Three level atom system in a cascade configuration. The transitions $a\to b$ and $b \to c$ are allowed, while $a \to c$ is forbidden. 
    The cavity field of frequency $\nu$ is detuned from $a\to b$ transition by an amount $\Delta = \nu - \omega_{ab}$.}
  \label{fig:system}
\end{figure}

Consider a three-level atom in a cascade configuration shown in Fig.~\ref{fig:system}. A $\pi/2$ pulse puts the atom into a coherent superposition
\begin{equation}
  \ket{\psi_{\rm A}}=\frac1{\sqrt2}(\ket b+\ket c),
\end{equation}
which passes through the cavity where it interacts with a single-mode electromagnetic field
\begin{equation}\label{before}
  \ket{\psi_{\rm L}}=\sum_{n=0}^\infty C_n\ket n.
\end{equation}
Here, $|C_n|^2$ is the probability of having $n$ photons, each  with a frequency $\nu$ detuned by $\Delta$ from
the $a-b$ transition frequency $\omega_{ab}$.
We assume that the detuning is large, i.e., $\frac{4g^2 n}{\Delta^2}\ll1$, where $n$ is the number of photons, and $g$ is the atom-light interaction strength. In this regime, the
population of the state $a$ is negligible, while the lower lying state $b$ attains a dispersive phase shift
\begin{equation}
  \ket b\rightarrow e^{-i n U_0 t}\ket b,
\end{equation}
where
$U_0=\frac{g^2}{\Delta}$,
with $t$ being the interaction time. At the output of the cavity, a $\pi/2$ pulse mixes the levels $b$ and $c$.
\begin{figure}[t!]
  \centering
  \resizebox{0.45\textwidth}{!}{%
   \includegraphics[]{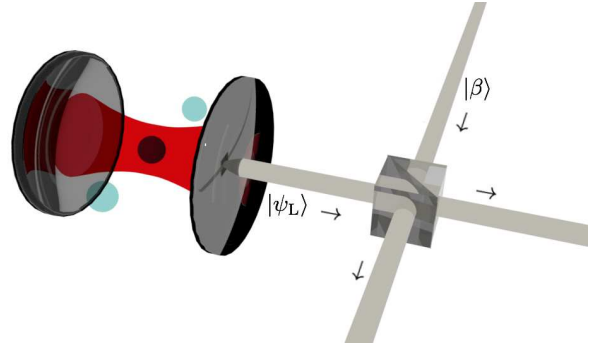}
}
  \caption{
    A schematic illustration of the interferometric setup. An atom passes a cavity where it interacts with a single mode of electromagnetic field. 
    After the atom leaves the cavity, its internal state is measured, and only one result is kept. In effect, a non-classical pulse of light $\ket{\psi_{\rm L}}$ is created and outcoupled from the cavity. 
    The beam mixes with a coherent state $\ket\beta$ on a first beam-splitter of the MZI.}
  \label{fig:setup}
\end{figure}
The result is an atom-light entangled state
\begin{align}
  \ket{\psi_{\rm A+L}}=& \sum_{n=0}^\infty C_n e^{-i\frac{nU_0t}2}\ket n   \nonumber \\ &\otimes\Big[\sin\left(\frac{nU_0t}2\right)\ket b+\cos\left(\frac{nU_0t}2\right)\ket c\Big].
\end{align}
Finally, the state of the atom is measured in the $b/c$
basis and only the outcomes where the atom is found in one of the states, say $\ket b$, are selected. Assuming that
initially the light is in a coherent state
with the amplitude $\alpha$, i.e.,
\begin{equation}\label{eq:coeff}
C_n=e^{-\frac{|\alpha|^2}2}\frac{\alpha^n}{\sqrt{n!}},
\end{equation}
we obtain that after the complete sequence the state of light is
\begin{equation}\label{after}
  \ket{\psi_{\rm L}}=\frac{1}{\sqrt{\mathcal A}}\sum_{n=0}^\infty C_n e^{-i\frac{nU_0t}2}\sin\left(\frac{nU_0t}2\right)\ket n,
\end{equation}
where $\mathcal A=\frac12\Big[1-e^{|\alpha|^2(\cos(U_0t)-1)}\cos\Big(|\alpha|^2\sin(U_0t)\Big)\Big]$ \cite{scully}.
Note that the state from Eq.~(\ref{after}) can be expressed as a sum of two coherent states, i.e.,
\begin{equation}\label{coh}
  \ket{\psi_{\rm L}}\propto\ket{\alpha\,e^{-iU_0t}}-\ket\alpha.
\end{equation}
Clearly, for $U_0t=\pi$, we obtain a superposition of two coherent states with opposite
amplitudes (Schr{\"o}dinger's cat state). In the following, we discuss an interferometric sequence utilizing the non-classical features of the state (\ref{coh}) for the SSN interferometry.

\section{Interferometric scheme and characterization}

We take the MZI with the port $a$ fed with $\ket{\psi_{\rm L}}$ from Eq.~(\ref{after}) and the port $b$ with a coherent state $\ket\beta$, namely
\begin{equation}\label{inp}
  \ket{\psi}=\ket{\psi_{\rm L}}_a\otimes\ket\beta_b.
\end{equation}
The MZI transfers the state (\ref{inp}) into
\begin{equation}\label{eq:interferometer}
  \ket{\psi(\theta)}=e^{-i\theta\hat J_y}\Big[\ket{\psi_{\rm L}}_a\otimes\ket\beta_b\Big],
\end{equation}
where $\hat J_y=\left(\hat a^\dagger\hat b-\hat b^\dagger\hat a\right)/2i$ is expressed in terms of the annihilation operators $\hat a$ and $\hat b$ for the two arms.
When the phase $\theta$ is estimated at the output ports of the interferometer, the sensitivity is bounded by the Cramer-Rao Lower Bound (CRLB)
\begin{equation}\label{crlb}
  \Delta\theta\geqslant\frac1{\sqrt m}\frac1{\sqrt{F_q}},
\end{equation}
where $F_q$ is called the quantum Fisher information (QFI) \cite{braunstein1994statistical}. The $F_q$ describes how much information about the parameter $\theta$ can be deposited in a quantum 
system,
and for a pure state transformed according to Eq.~(\ref{eq:interferometer}), it takes a particularly simple form
\begin{equation}\label{qfi}
  F_q=4\left[\av{\hat J_y^2}-\av{\hat J_y}^2\right],
\end{equation}
where the mean values are calculated using the output state (\ref{eq:interferometer}). 
Equation (\ref{crlb}) tells that the higher the value of the QFI, the better the sensitivity can be achieved. 
\begin{figure}[t!]
  \centering
    \resizebox{0.45\textwidth}{!}{%
  \begin{tabular}{cc}
    \vspace*{-1.2cm} \includegraphics[clip,scale=0.9]{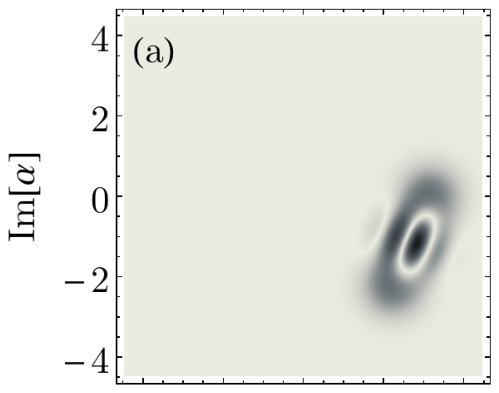} & \includegraphics[trim=0.3cm -0.76cm 0 0 clip,scale=0.87]{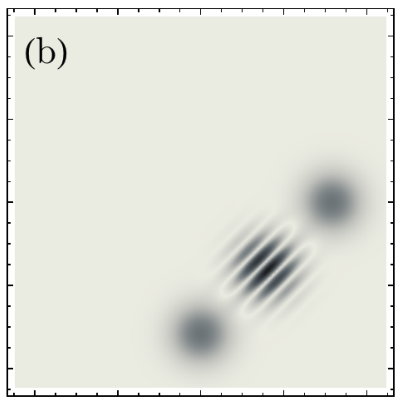} \\
    \includegraphics[trim=-1.5mm -0.76cm 0 0 clip,scale=0.905]{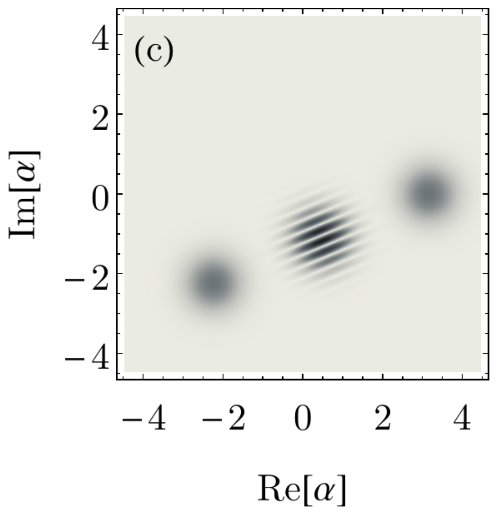} & \includegraphics[trim=0.3cm -0.83cm 0 0 clip,scale=0.869]{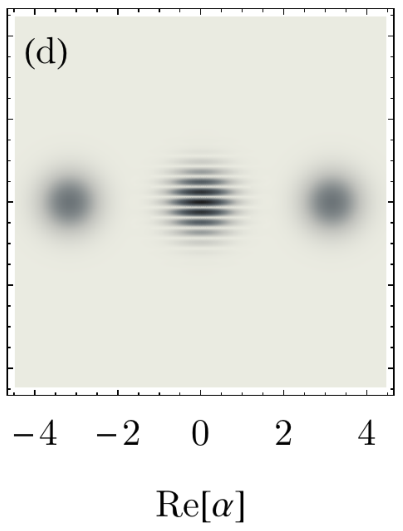}
  \end{tabular} 
}
  \caption{Wigner function of the state (\ref{after}) displayed at $U_0 t=\pi/4$ {\bf(a)}, $U_0 t=\pi/2$ {\bf(b)}, $U_0 t=3\pi/4$ {\bf(c)} and $U_0 t=\pi$ {\bf(d)}. 
    Directions perpendicular to fringes indicate the direction of optimal interferometer.}
  \label{fig:exp}
\end{figure}

At every instant of time, the value of the $F_q$ can be optimized by a proper choice of the complex amplitude $\beta=\beta_0e^{i\varphi_\beta}$ of the coherent beam. To read out which 
phase $\varphi_\beta$ is optimal (with the amplitude $\beta_0$ fixed), we first note that
the precision of the phase estimation is directly
linked to  the distinguishability of the neighboring states $\ket{\psi(\theta)}$ and $\ket{\psi(\theta+\delta\theta)}$, i.e., 
\begin{equation}\label{dist}
  |\bra{\psi(\theta)}\psi(\theta+\delta\theta) \rangle|^2 \simeq1- \frac{(\delta\theta)^2}{8}F_q.
\end{equation}
According to this formula, the higher the $F_q$, the more the two states differ, and in consequence, the parameter $\theta$ can be estimated with a higher resolution 
\cite{braunstein1994statistical,wootters}, see Eq.~(\ref{crlb}). 
On the other hand, the same quantity (\ref{dist}) can be approximately
written using the Wigner function of the state $\ket{\psi_{\rm L}}$ defined as
\begin{equation}
  \mathcal{W}(\alpha)=\int\! \frac{\rm{d}^2\lambda}{\pi^2}\ \bra{\psi_{\rm L}}e^{\lambda(\hat a^{\dagger}-\alpha^*)-\lambda^*(\hat a-\alpha)}\ket{\psi_{\rm L}}.
\end{equation}
If the intensity of the coherent beam is high, the MZI transformation can be approximated by replacing the annihilation and the creation operators for the mode $b$ with the complex
numbers $\beta$ and $\beta^*$, i.e.,
\begin{equation}\label{eq:wignerdis}
  e^{-i\theta\hat J_y}=e^{-\frac\theta2\left(\hat a^\dagger\hat b-\hat b^\dagger\hat a\right)}\approx e^{-\frac\theta2\left(\hat a^\dagger\beta-\beta^*\hat a\right)}.
\end{equation}
This is the displacement operator which shifts the Wigner function in the complex plane by the distance $\frac{\beta_0\theta}2$ in the direction set by the phase $\varphi_\beta$. To complete the picture,
we notice that the scalar product from Eq.~(\ref{dist}) can be expressed in terms of the overlap of the Wigner functions
\begin{equation}\label{eq:fidfish}
  |\bra {\psi (\theta)}\psi(\theta+\delta\theta) \rangle|^2=\pi \int\!\!\rm{d}^2\alpha\, \mathcal W(\alpha) \mathcal W(\alpha+\delta\alpha),
\end{equation}
where $\delta\alpha=\frac{\beta_0\cdot\delta\theta}2$. By combining Equations (\ref{dist}) and (\ref{eq:fidfish}), we conclude that high values of the QFI require low
overlap of the Wigner functions of the neighboring states. 

We now plot $\mathcal W$ for four different instants of time $U_0t=\frac\pi4,\frac\pi2,\frac{3\pi}4,\pi$, see Fig.~\ref{fig:exp}. The emergence of the interference fringes at later times
signals the growing quantumness of the state, and to obtain a minimal overlap one should shift the Wigner function in the direction where the change (i.e., the gradient) of $\mathcal W$ is
maximal. According to Eq.~(\ref{eq:wignerdis}) this condition determines the phase $\varphi_\beta$ of the coherent beam. For instance, the panel (b) of Fig.~\ref{fig:exp} indicates that
the phase should be $\varphi_\beta=-\frac\pi4$, while for (d) it should be $\varphi_\beta=\pi$ to make a shift along the imaginary axis. 

\begin{figure*}[t!]
  \centering
     \resizebox{0.9\textwidth}{!}{%
  \includegraphics[clip,scale=0.85]{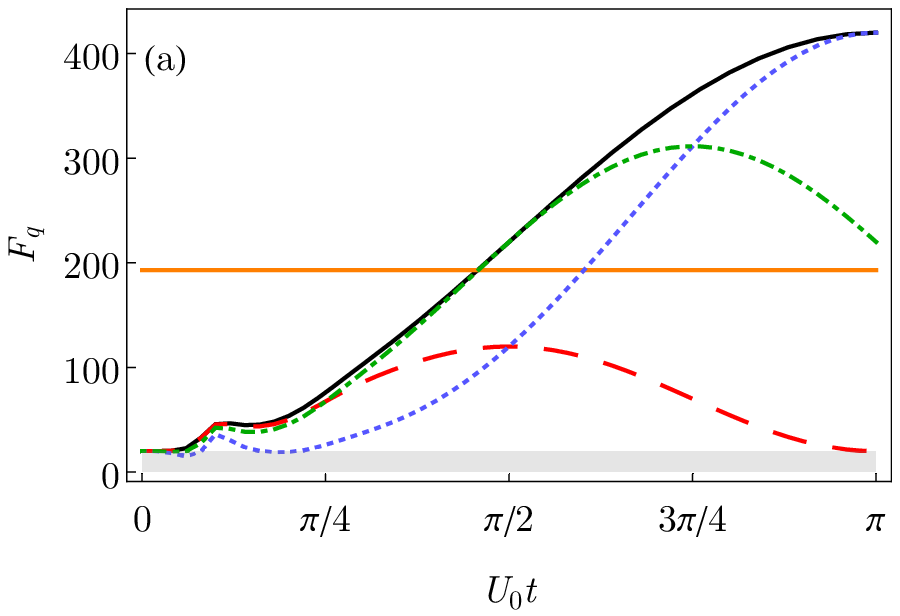}
  \includegraphics[clip,scale=0.85]{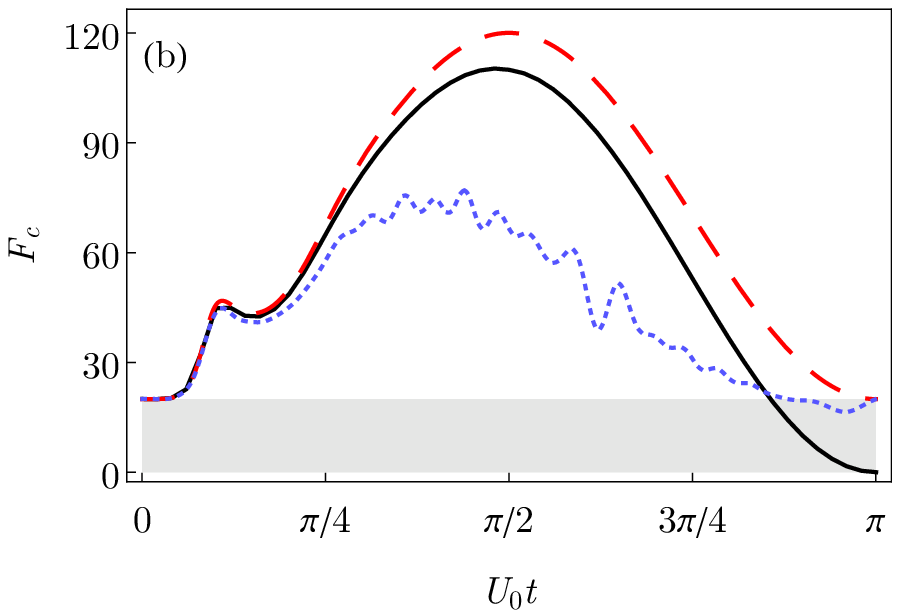}
  }
  \caption{{\bf (a)} Quantum Fisher information as a function of the rescaled cavity-atom interaction time $U_0 t$ for different phases of the coherent amplitude of the field $\ket\beta$.
    The solid line is optimized over the phase $\varphi_\beta$ at every instant. Also, we show the QFI for $\varphi_\beta=\pi/2$ (dotted blue), $\varphi_\beta=\pi/4$ (dot-dashed green) and 
    $\varphi_\beta=0$ (dashed red). The solid orange line is the limit one could 
    achieve with 10.3 dB squeezed vacuum field (see Ref \cite{aasi2013enhanced}) on one port and a coherent state 
    on the other one given $n_{\alpha}+n_{\beta}=20.$
    {\bf (b)} $\varphi_\beta=0$ case: the $F_c$ calculated with Eq.~(\ref{class}) at $\theta=0$ (solid black) and  $\theta=\pi/13$ (dotted blue), compared with the QFI (dashed red).
    All results for $n_{\alpha/\beta}=10$. Here and in the following figures, the gray area represents the classical interferometry regime.}
  \label{fig:qfi}
\end{figure*}

When the overlap between the two components of the state (\ref{coh}) is negligible, the $F_q$ optimized at every instant with respect to the phase $\varphi_\beta$ reads
\begin{equation}\label{eq:approxqfi}
  F^{\rm opt}_q\approx n_\alpha+n_\beta+4n_\alpha n_\beta\sin^2(U_0t),
\end{equation}
where $n_{\alpha/\beta}$ is the mean number of photons in each beam.
On the other hand, if one keeps $\varphi_\beta=0$ for all times, the QFI is well approximated by the formula
\begin{equation}\label{ph0}
  F^{\varphi_\beta=0}_q\approx n_\alpha+n_\beta+n_\alpha n_\beta\sin^2\left(\frac{U_0t}2\right).
\end{equation}
Figure~\ref{fig:qfi}(a) shows  the QFI calculated with different choices of $\varphi_\beta$ where for the reference we also show the limit that one could achieve with a squeezed vacuum state \cite{aasi2013enhanced}. The gain from the Wigner-function approach is that we now
intuitively understand what is  a proper choice of the coherent state $\ket\beta$ at every moment of the evolution.

From now on we fix $\varphi_\beta=0$ and take $\alpha\in\mathbb{R}$ in all the calculations. 
This is not always an optimal choice, and the comparison of Eqs (\ref{eq:approxqfi}) with (\ref{ph0}) reveals a factor of four deterioration
of the QFI in the maximal value, nevertheless, the purely real coherent amplitude $\beta$ retains the Heisenberg scaling. 

Once the coherent field is fixed and the ultimate sensitivity provided by the CRLB from Eq.~(\ref{crlb}) is evaluated through the Eq.~(\ref{ph0}), we proceed to calculate the 
sensitivity in a particular estimation protocol based  on the
measurement of the number of photons $n$ and $m$ in the output
ports. With the probability
\begin{equation}\label{eq:probability}
  p_{nm}(\theta)=\Big|\braket{n,m}{\psi(\theta)}\Big|^2
\end{equation}
at hand, we use the maximum
likelihood estimator to deduce the value of $\theta$ \cite{holevo2011probabilistic}. 
The sensitivity in such case is bounded by the CRLB (\ref{crlb}) with the QFI $F_q$ replaced by
\begin{equation}\label{class}
  F_c=\sum_{n,m}\frac{1}{p_{nm}(\theta)}\left(\partial_\theta p_{nm}(\theta)\right)^2
\end{equation}
In Fig.~\ref{fig:qfi}(b), we display the $F_c$ calculated with $n_{\alpha/\beta}=10$ photons as a function of the interaction time for two different values of the phase. 
We observe a strong dependence of the FI on $\theta$, however the numerical tests for different $\alpha$'s and $\beta$'s reveal that at $\theta=0$ the 
FI is smaller than the QFI by only $(n_\alpha+n_\beta)/2$ and retains the Heisenberg scaling.

\begin{figure}[t!]
\centering
   \resizebox{0.45\textwidth}{!}{%
  \begin{tabular}{cc}
 \includegraphics[clip,scale=0.9]{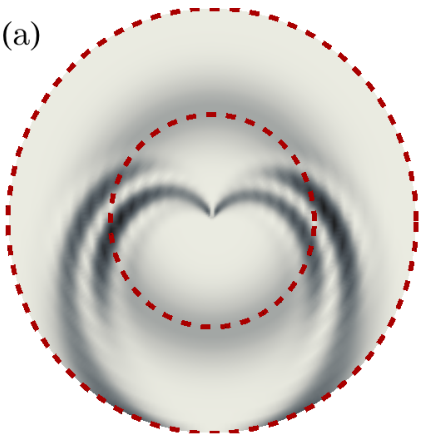} & \includegraphics[clip,scale=0.9]{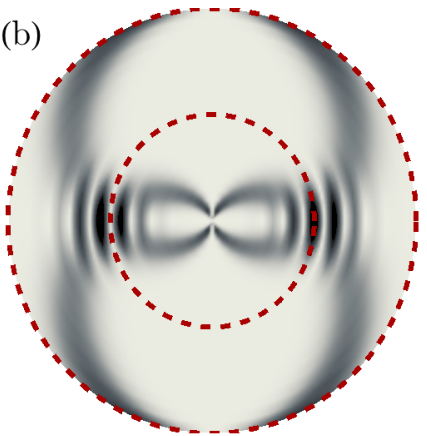} \\
    \includegraphics[clip,scale=0.9]{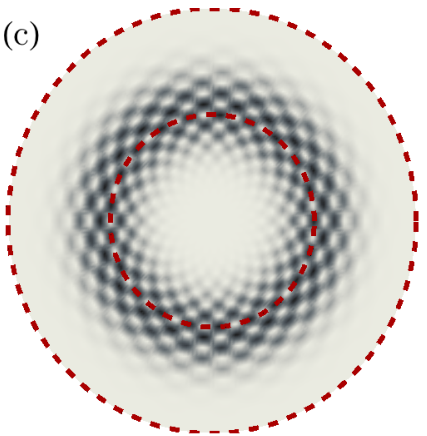} & \includegraphics[clip,scale=0.9]{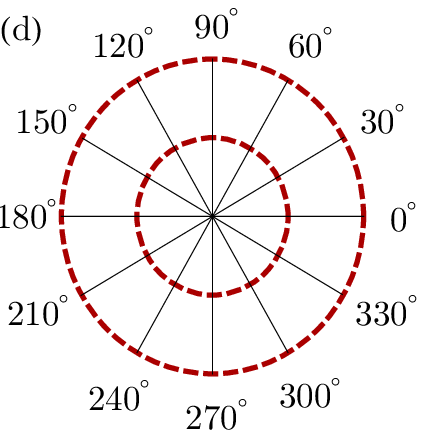}
  \end{tabular} 
}
\caption{The probability of measuring a fixed number of photons ($N=20$) as a
  function of $\theta$ (polar variable) and relative photon number $\Delta N$ (radial
  variable). {\bf(a)} and {\bf(b)} correspond to real $\beta$ with  $U_0 t=\pi/2$ and  $U_0 t=\pi$. {\bf(c)} 
  used imaginary $\beta$ and  $U_0 t=\pi$, {\bf(d)} is the polar grid. 
  Darker regions correspond to larger probability.}
\label{fig:exp4}
\end{figure}

We now further inspect the strong dependence of the FI on $\theta$. To this end, we pick a subspace of a fixed number of photons $n+m=40$ and plot the probability from Eq.~(\ref{eq:probability}) 
as a function of $\theta$ and the relative number of photons $\Delta n=n-m$.
Figure \ref{fig:exp4} reveals the presence of fine structures in the probability, which translate to high 
interferometric sensitivity \cite{PhysRevA.92.043622,Wasak2016}. 

\section{Impact of imperfections}

\subsection{Cavity losses}

We now examine the impact of the photon losses from the cavity during the state-preparation. 
To this end, we incorporate a Lindblad term into the Heisenberg equation for the atom-light density matrix
in the dispersive regime, namely
\begin{align}\label{melf}
  \partial_t\hat{\varrho} &= - \frac{i}{\hbar}\left[\hat{H},\hat{\varrho}\right]+\mathcal{L}[\hat{\varrho}] \nonumber \\&=-\frac{ig^2}{\hbar\Delta}\Big[ \hat{a}^{\dagger}\hat{a} \ket b \bra b,\hat{\varrho}\Big]
  +\kappa\left(\{\hat a^{\dagger} \hat a,\hat{\varrho}\}-2 \hat a \hat{\varrho} \hat a^{\dagger}\right),
\end{align}
where $\kappa$ is the loss rate.
Just as previously, after the atom leaves the cavity a $\pi/2$-pulse mixes the two internal states $\ket b$ and $\ket c$, 
and subsequently the state of the atom is measured with only the $\ket
b$ results kept. As a result of this sequence, due to decoherence a mixed state of light $\hat{\varrho}_{\rm L}(t)$ is generated, 
contrary to the ideal case of Eq.~(\ref{after}).
\begin{figure*}[t!]
  \centering
     \resizebox{0.9\textwidth}{!}{%
  \includegraphics[clip,scale=0.815]{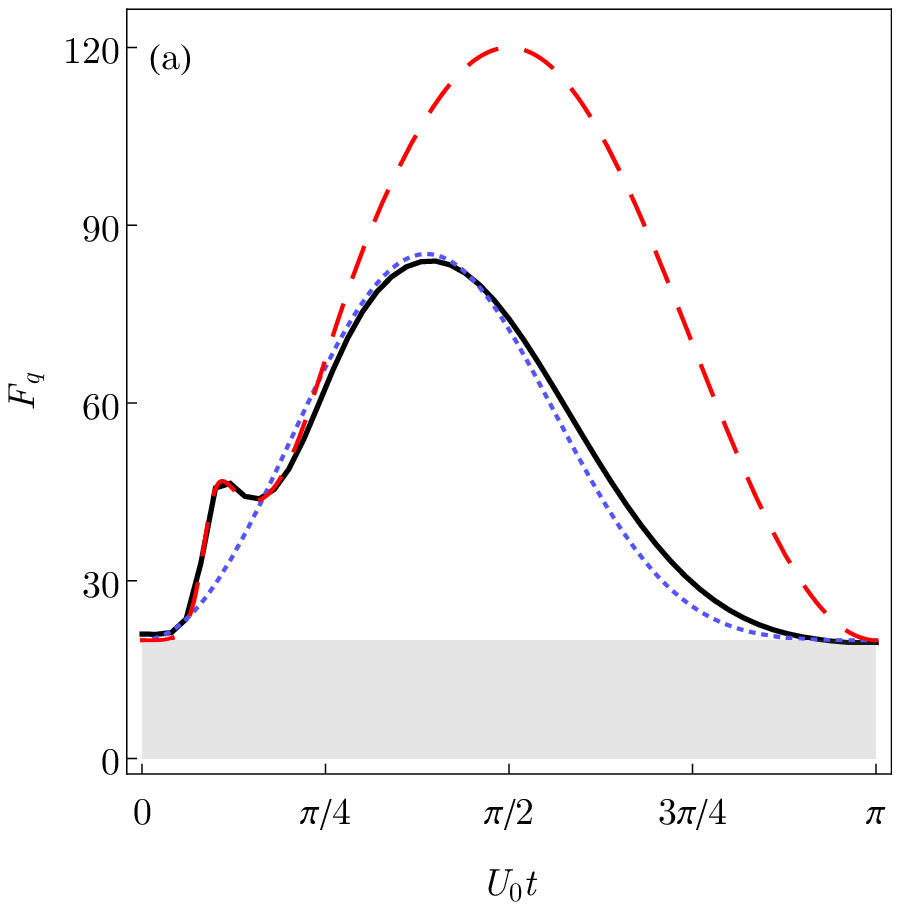}
  \includegraphics[clip,scale=.8]{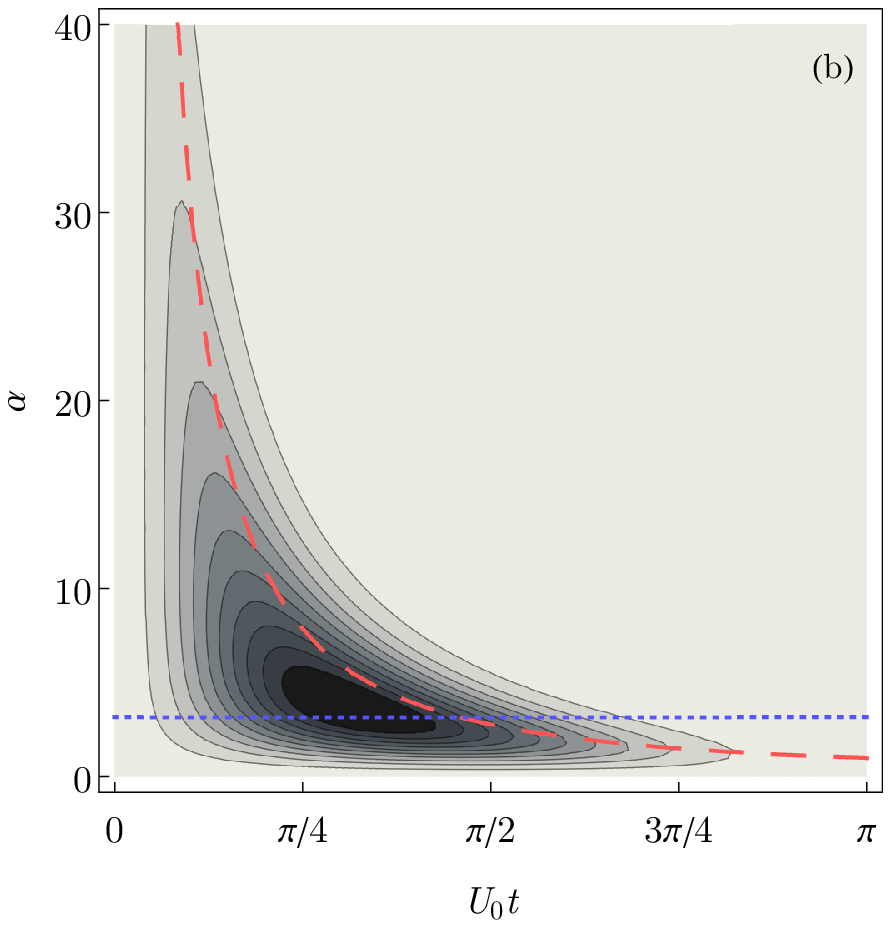}
  }
  \caption{{\bf (a)}  Behavior of the QFI in presence of cavity losses during state preparation for $n_\alpha=10$, $n_\beta=10$, $\kappa=0.05$,
    and $U_0=2$. The solid black line is the exact QFI calculated with Eq. (\ref{qfifull}) and the blue dotted line is the QFI calculated with the approximated formula Eq. (\ref{fql}).
    The dashed red line is the QFI for $\kappa=0$. 
    {\bf (b)} The QFI as a function of $\alpha$ and
    the interaction time. 
    The dashed-red line represents the condition $t=\tau$, and the dotted-blue line is the cut shown in {\bf(a)}. }
  \label{fig:compar}
\end{figure*}

First, we characterize the performance of an interferometer in the presence of losses with the QFI which for mixed states is
\begin{equation}\label{qfifull}
  F_q = 2 \sum_{i\neq j} \frac{(\lambda_i-\lambda_j)^2}{\lambda_i+\lambda_j}|\bra i \hat{J}_y \ket j |^2,
\end{equation}
where $\lambda_{i/j}$ and $\ket{i/j}$ are the eigenvalues and the
eigenvectors of the density matrix propagated through the MZI
 \cite{braunstein1994statistical}, i.e., 
\begin{equation}\label{den_mzi}
  \hat\varrho(\theta)=e^{i\theta\hat J_y}\left[\hat{\varrho}_{\rm L}(t)\otimes\ket\beta\bra\beta\right]e^{-i\theta\hat J_y}.
\end{equation} 
Although analytical predictions similar to those from Eq.~(\ref{eq:approxqfi}) are not possible anymore, numerical inquiry reveals the presence
of a time scale $\tau$ associated with the strength of losses and the interactions, $\kappa \tau (U_0\tau)^2n_\alpha=1$.
When $t\lesssim\tau$ the numerical results for the QFI are well fitted
with
\begin{equation}\label{fql}
  F_{q}\approx n_\alpha+n_\beta+\exp\left[-\frac{2}{3}\left(\frac{t}{\tau}\right)^3\right]\sin^2(U_ot)n_\alpha n_\beta.
\end{equation}
This simple phenomenological formula illustrates how the nonlinear term responsible for the SSN scaling is suppressed by cavity losses. As the time 
grows,
the suppression is more significant. Also, when the intensity $n_\alpha$  is high, the losses are more harmful.

A comparison between the approximate expression from Eq.~(\ref{fql})
and the complete calculation (\ref{qfifull}) is shown in Fig.~\ref{fig:compar}(a).
Figure \ref{fig:compar}(b) shows in addition that increasing $n_\alpha$ does not increase the QFI. Clearly, for a given set of parameters, there is an optimal choice 
of the interaction inside the cavity and the intensity $n_\alpha$. The formula (\ref{fql}) allows for a rough estimate of the working point of the interferometer. 
\begin{figure*}[t!]
  \centering
     \resizebox{0.9\textwidth}{!}{%
  \includegraphics[clip,scale=0.83]{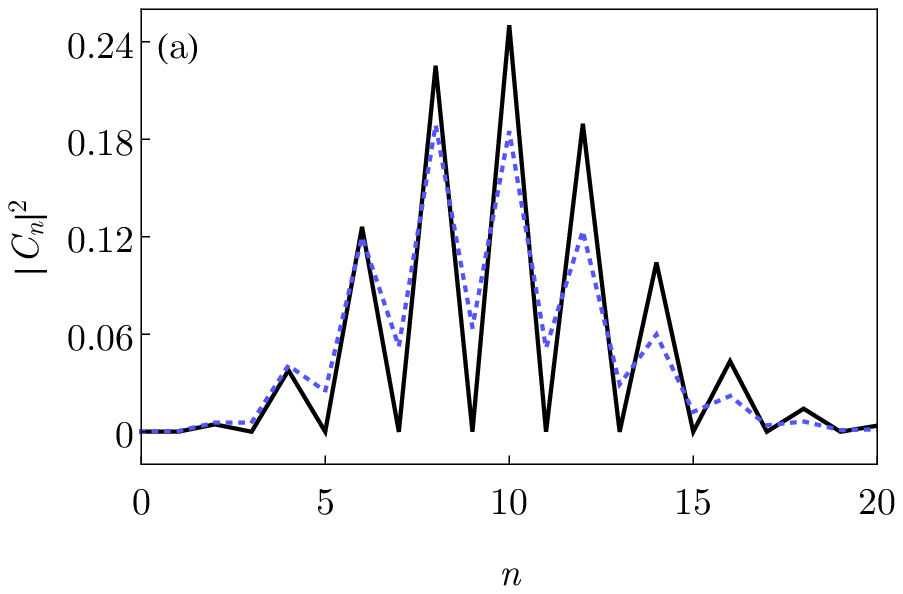}
  \includegraphics[clip,scale=0.8]{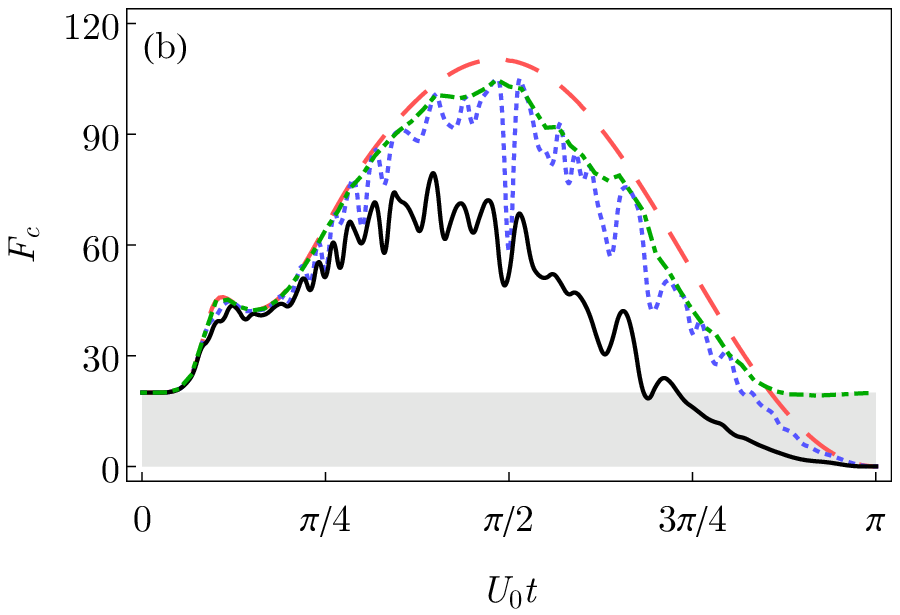}
  }
  \caption{{\bf (a)} Effect of the photon losses on the number distribution of the light field with initally $n_\alpha=10$, outgoing from the cavity at $U_0t=\pi$ for an ideal case (solid black line)
    and $\kappa=10^{-2}$ case (dotted blue line).
    {\bf (b)}
    The FI calculated at $\theta=0$ with $n_{\alpha/\beta}=10$ for: $\kappa=0$ (dashed red), $\kappa=10^{-3}$ (dotted blue) and $\kappa=10^{-2}$ (solid black). 
    Dot-dashed green line is the $\kappa=10^{-3}$ case optimized over $\theta$ for every instant.}
\label{fig:lindblad}
\end{figure*}

\begin{figure}[t!]
  \centering
       \resizebox{0.45\textwidth}{!}{%
  \includegraphics[clip,scale=0.7]{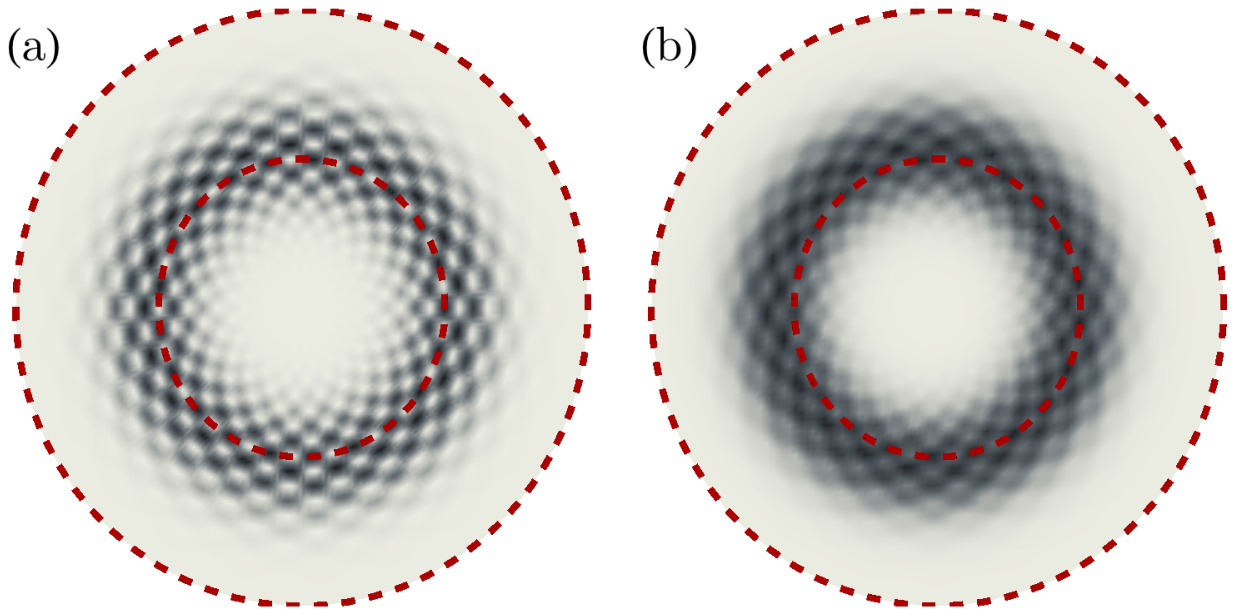}
  }
\caption{ The probability of measuring $N=20$ photons as a
    function of $\theta$ (polar variable) and relative photon number $\Delta N$ (radial
    variable). Panels depitc state (\ref{coh}) with $U_0 t=\pi$ for an ideal case {\bf(a)} and for a lossy case with $\kappa=10^{-2}$ 
    {\bf(b)}.}
  \label{fig:exp3}
\end{figure}

Knowing the lower bound for the sensitivity, we focus again on the estimation from the measurement of the number of photons.
First, we plot the photon number distribution of the signal outgoing from the cavity. In the absence of losses, the interaction of the cavity field with the atom imprints oscillations
in this distribution (see the sine function in Eq.~(\ref{after})). These fine structures, drawn for the no-loss case
with a solid black line in Fig.~\ref{fig:lindblad}(a) drive the high performance of this particular estimation scheme \cite{pezze2009entanglement,wasak2015interferometry,PhysRevA.92.043622}.
Even moderate losses smooth out these structures, as depicted by the dotted blue line and shown in more detail in Fig.~\ref{fig:exp3}. This indeed has a profound impact on the sensitivity, as 
confirmed by the numerical calculations of the Fisher information shown in Fig.~\ref{fig:lindblad}(b) for $\kappa=0,\,0.001$ and $0.01$. 
For each case, the FI is calculated with the formula (\ref{class}) where the probability from Eq.~(\ref{eq:probability}) is generalized to mixed-states
\begin{equation}\label{ideal}
  p_{nm}(\theta)=\mathrm{Tr} \left[\ket{n,m}\bra{n,m}\hat\varrho(\theta)\right]
\end{equation}
and $\hat\varrho(\theta)$ is defined in Eq.~(\ref{den_mzi}). 
We notice that not only the $F_c$ decreases, but also as soon as 
$\kappa\neq0$, it reveals fast oscillations as a function of the interaction time. The value of the FI varies strongly as $\theta$ is changed, though this dependence can be smoothed out by
choosing an optimal $\theta$ at each time, as shown by the green line in Fig.~\ref{fig:lindblad}(b).

\subsection{Imperfect photon counting}

The other imperfection we consider is the finite resolution of the photon counting at the output ports of the MZI.
We model this effect by the convolution of the ideal-case probability from Eq.~(\ref{ideal}) of measuring $n'$ and $m'$ photons with a Gaussian distribution
\begin{equation}
  \mathcal{P}(n,m|\theta)=\mathcal{N}\sum_{n',m'=0}^{\infty}{p}(n,m|\theta)e^{-\frac{(n' - n)^2+(m' - m)^2}{2 \sigma^2}},
\end{equation}
where $\sigma$ accounts for the level of uncertainty, and $\mathcal{N}$ is the normalization constant.

\begin{figure}[t!]
  \centering
       \resizebox{0.45\textwidth}{!}{%
  \includegraphics[clip,scale=1.0]{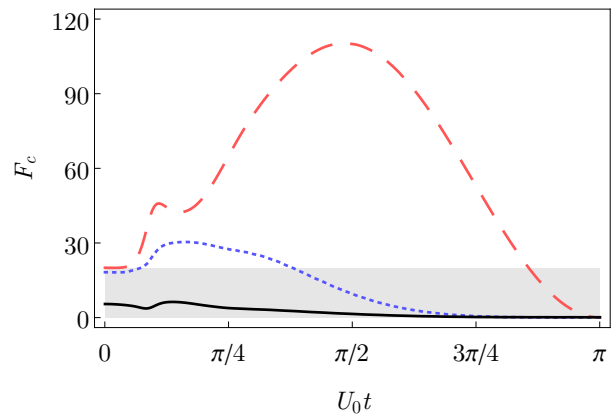}
  }
  \caption{Fisher information with imperfect detectors for $n_{\alpha/\beta}=10$ and $\sigma=0$ (dashed red), $\sigma=1$ 
    (dotted blue) and $\sigma=5$ (solid black).}
  \label{fig:regime}
\end{figure}

We display the FI calculated with three different values of $\sigma=0,\,1$ and 5, see Fig.~\ref{fig:regime}. Clearly, imperfections in the photon counting have a significant impact
on the sensitivity. Characteristically for any implementation of non-classical states, improvement beyond the SNL requires highly efficient detectors.

\section{Implementation}

The scheme for tailored creation of entangled states of a light mode and a single atom was proposed decades ago \cite{phoenix1991establishment} and experimentally first implemented at 
ENS Paris using the superconducting microwave cavities \cite{haroche_rmp,rauschenbeutel1999coherent}. The subsequent measurement of the atomic population allows to implement 
it as the QND measurement of the light field, giving highly non-classical  states \cite{rauschenbeutel2001controlled}.
These ideas were extended to the optical regime
\cite{boozer2007reversible,lettner2011remote} where the long-distance propagation of photons can be readily
exploited to implement basic quantum information processing tasks \cite{kimble2008quantum}. 
Using excitation sequences, tailored superposition of propagation photon states were engineered with a single atom in a cavity \cite{weber2009photon} 
as well as controlled nonlinear phase shifts \cite{volz2014nonlinear}.

As we focus here on the interferometric applications, the optical regime where photons can be efficiently 
extracted from the cavity and recombined on a beam-splitter 
is the operating regime of choice, though the physics for microwaves does not differ.
The need to extract photons from the cavity directly reveals the twofold role of the photon leakage. 
On the one hand, during the state preparation photon leakage is detrimental to the atom-field 
entanglement as it provides potential information on the state of the system \cite{ritsch2004deterministic} 
and therefore limits the interferometric precision. At first sight this suggests fast state preparation 
during which photons are not detected outside the cavity.

On the other hand, the non-classical field inside the cavity is transformed to entangled photons solely 
via leakage through the mirrors, which acts as beam splitters \cite{asboth2005computable}. 
Clearly, it is crucial to extract enough photons out of the mode to create a propagating non-classical wave-packet.
For cat states (achieved in the ideal preparation scheme), a single photon lost from the wave-packet 
renders the output state classical \cite{ritsch_2004}.
However, as we will show next, the SNL can still be beaten in
presence of an imperfect extraction once we are dealing with more classical states, 
as those obtained after the imperfect preparation scheme discussed
above.

\begin{figure}[t!]
  \centering
       \resizebox{0.45\textwidth}{!}{%
  \includegraphics[clip,scale=1.0]{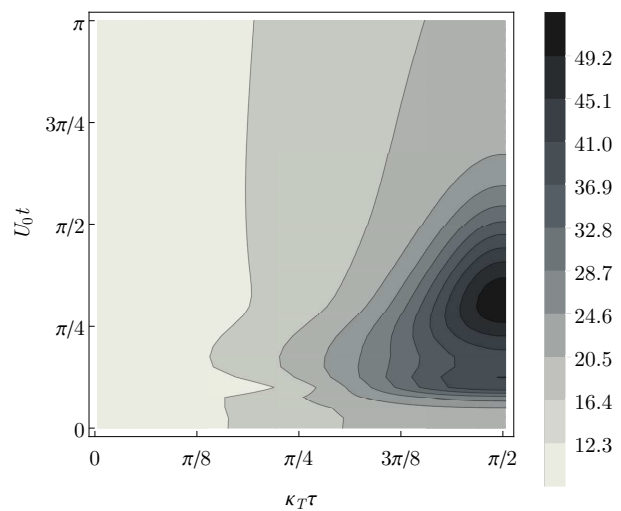}
  }
  \caption{QFI for Mach-Zehnder interferometer as a function of interaction and transfer time. 
  QFI is calculated by taking the light extracted from the cavity on one of the interferometr's input port 
  and a coherent field on the other one. Here $\kappa_T=0.049$ and $\tilde{\kappa}=0.001$. 
  At transfer time equal to ${\pi}/{2 \kappa_T}$ all the photons are extracted from the cavity, and at interaction time $\pi/U_0$ the cat state is created in the cavity.}
  \label{fig:transfer}
\end{figure}

In order to examine how the imperfect extraction of photons from the cavity 
affects QFI, we propose a simple, albeit qualitative, model in which the extracted photons are treated as a new mode $\hat b$. 
In this picture, one of the mirrors in the cavity serves as a beam-splitter
which can absorb photons, and thus the time evolution of the density matrix is
\begin{align}
  \partial_t \hat \varrho &= - \frac{i}{\hbar}\left[\hat H_T, \hat \varrho\right] + 
  \mathcal{L}\left[\hat \varrho\right] \nonumber\\
  &= -\frac{i \kappa_T}{ \hbar} \left[\hat a^\dagger\hat b+ \hat b^\dagger \hat a, \hat \varrho\right] 
  + \tilde{\kappa}\left(\{\hat a^\dagger \hat a, \hat  \varrho \}-2 \hat a \hat \varrho \hat 
  a^\dagger\right),
\end{align}
where $\kappa_T$ is the rate at which the photons are extracted from the cavity, 
and $\tilde{\kappa}$ is the loss rate corresponding to photons being
absorbed within the mirror and therefore not reaching the input of the
interferometer. The behavior of the QFI as a function of the
preparation time $t$ and the extraction time $\tau$ is shown in
Fig.~\ref{fig:transfer} for a fixed value of the mirror absorption
rate $\tilde{\kappa}$.
As expected, the QFI increases as 
a function of transfer time up to the moment when all the photons are 
transferred: $\kappa_T\tau=\pi/2$. For $\tau=0$ the cavity output does
not reach the interferometer, which is then fed with a single coherent
state, resulting into a value of QFI corresponding to half of the SNL
(compare with Fig.~\ref{fig:compar}(a)). For $\kappa_T\tau=\pi/2$,
there is an optimal atom-cavity interaction time $U_0t$ for the largest
QFI to be achieved, the latter depending obviously on the absorption
rate $\kappa$.
Comparison with Fig.~\ref{fig:compar}(a) reveals i) that the optimal
interaction time is decreased to a finite absorption rate, ii) the
largest QFI at the optimal time is also decreased as a result of the
additional decoherence channel. As
anticipated, the SNL can still be overcome by using more classical
intracavity states.

As already discussed, the transfer rate $\kappa_T$ and the cavity loss
rate $\kappa$ affecting the state preparation are not independent in
general and actually essentially of the same order in a standard
Fabry-Perot cavity. It is thus clear how this situation requires an
optimisation of the coupled quantities $\kappa$ and $\kappa_T$.
To avoid such timing problems, an ideal experimental setup would
include the possibility for fast control of the photon loss rate, keeping it as small as possible
during the preparation stage, while switching it to a much higher value after the state is prepared \cite{reinhard2012strongly}. 
While technically not easy, the cavity decay and coupling can be quickly tuned in nano-fiber setups with evanescent wave coupling. 
This simultaneously minimizes unwanted mirror losses allowing for optimized photon out-coupling and routing \cite{aoki2009efficient}.

\section{Conclusions}

We have studied a Mach-Zehnder interferometer operating with the highly non-classical light generated by nonlinear atom-light interaction in a high-$Q$ Cavity QED system. An injected coherent light pulse gets dynamically entangled with the internal atomic state, so that a subsequent projective measurement of the latter yields a highly non-classical state of light, conditioned on the measurement outcome.
When this state is injected into one port of the MZI in conjunction
with coherent light at the other port, the system exhibits a strongly
enhanced phase sensitivity significantly surpassing the SNL. To test
and benchmark a practical measurement procedure, we suggest and
numerically evaluate an efficient phase shift estimation scheme based
on number resolved counting of photons at the interferometer
output. Photon losses and photon counting errors deteriorate the
interferometer sensitivity, but it proves to still be significantly
better than the shot-noise limit under realistic conditions.

\section{Acknowledgements}
  
KG acknowledges the support of the Erasmus Mundus programme of the European Union. FP acknowledges the support by the APART fellowship of the Austrian Academy of Sciences. HR is supported by the Austrian Science Fund project I1697- N27.
This work was supported by the Polish Ministry of Science and Higher Education programme ``Iuventus Plus'' (2015-2017) Project No.~IP2014~050073, 
and the National Science Center Grant No. 2014/14/M/ST2/00015.

\end{document}